\documentclass[aps,pra,reprint,superscriptaddress,longbibliography]{revtex4-1}

\usepackage{amsmath}    
\usepackage{graphicx}   
\usepackage[lofdepth,lotdepth, caption=false]{subfig}
\usepackage{verbatim}   
\usepackage{color}      
\usepackage{hyperref}   
\usepackage{dcolumn}
\usepackage{bm}
\usepackage{miller}

\usepackage{color}

\def\lsco{La$_{2-x}$Sr$_x$CuO$_4$}
\def\lbco{La$_{2-x}$Ba$_x$CuO$_4$}

\def\ybco{YBa$_2$Cu$_3$O$_{6+x}$}

\def\bscco{Bi$_2$Sr$_2$CaCu$_2$O$_{8+\delta}$}

\def\lco{La$_2$CuO$_4$}
\newcommand{\lcco}{La$_{2-x}$Ca$_{1+x}$Cu$_{2}$O$_{6}$}

\newcommand{\degrees}{$^\circ$C}

\def\newr{\color{black}}

\begin{document}

\title{Gapless spin excitations in superconducting La$_{2-x}$Ca$_{1+x}$Cu$_{2}$O$_{6}$ with $T_c$ up to 55~K}
\author{John A.~Schneeloch}
\thanks{Present address: Department of Physics, University of Virginia, Charlottesville, Virginia 22904, USA}
\affiliation{Condensed Matter Physics and Materials Science Division, Brookhaven National Laboratory, Upton, New York 11973, USA}
\affiliation{Department of Physics and Astronomy, Stony Brook University, Stony Brook, New York 11794, USA}

\author{Ruidan Zhong}
\thanks{Present address: Department of Chemistry, Princeton University,
Princeton, New Jersey 08544, USA}
\affiliation{Condensed Matter Physics and Materials Science Division, Brookhaven National Laboratory, Upton, New York 11973, USA}
\affiliation{Materials Science and Engineering Department, Stony Brook University, Stony Brook, New York 11794, USA}

\author{M. B.~Stone}
\affiliation{Neutron Scattering Division, Oak Ridge National Laboratory, Oak Ridge, Tennessee 37831, USA}

\author{I. A. Zaliznyak}
\author{G. D. Gu}
\author{Guangyong Xu}
\thanks{Present address: NIST Center for Neutron Research, National Institute of Standards and Technology, 100 Bureau Drive, Gaithersburg, Maryland, 20899, USA}
\author{J.~M.~Tranquada}
\affiliation{Condensed Matter Physics and Materials Science Division, Brookhaven National Laboratory, Upton, New York 11973, USA}

\date{\today}

\begin{abstract}
We report inelastic neutron scattering on single crystals of the bilayer cuprate family La$_{2-x}$Ca$_{1+x}$Cu$_{2}$O$_{6+\delta}$, including two crystals made superconducting (transitions at 45 and 55~K) by high-pressure annealing in an oxygen-containing atmosphere. The magnetic excitations in the non-superconducting crystal have a similar temperature-dependence as those in weakly hole-doped cuprates. In the superconducting crystals, there is a near-uniform suppression of the magnetic spectral weight with increasing temperature; in particular, there are no signs of a spin gap or ``resonance'' peak. This is different from the temperature dependence seen in many optimally-doped cuprates but similar to the behavior seen in certain underdoped cuprates.   We discuss the possible connection with pair-density-wave superconductivity.

\end{abstract}

\maketitle

\section{Introduction}

While the mechanism of superconductivity in the high-$T_{c}$ cuprate superconductors remains controversial, inelastic neutron scattering measurements of magnetic correlations may give insight into excitations widely believed to play an important role in cuprate superconductivity. The cuprates consist of many families, each having CuO$_{2}$ planes and a similar phase diagram \cite{keim15}, but each also having quirks such as differing numbers of CuO$_{2}$ planes per unit cell, interlayer structures, type and mobility of dopants, etc.  It is important to conduct measurements on many different cuprate families to distinguish between properties universal to all cuprates or particular to certain families.  One family infrequently studied by neutrons is the La$_{2-x}$(Ca,Sr)$_{x}$CaCu$_{2}$O$_{6+\delta}$ (La-2126) family \cite{cava90,fuer90,kino90,liu91,kino92b}, which is a bilayer variant of the more commonly studied ``La-214'' family having La$_{2}$CuO$_{4}$ as the parent compound.  

Synthesis of large, superconducting La-2126 crystals is difficult, as substitution of Ca or Sr for La at sufficient levels to yield superconductivity requires annealing in a high-pressure oxygen-containing atmosphere \cite{ishi91,okuy94}, and such facilities are relatively uncommon.  We have recently succeeded in growing and oxygen-annealing large crystals of Ca-doped La-2126, achieving superconducting transition temperatures, $T_c$, as high as 55~K \cite{shi17,schn17,gu06a}.  In contrast to La-214, the La-2126 structure contains CuO$_2$ bilayers.  A transport study on La-2126 crystals demonstrated decoupling of the superconducting bilayers in a magnetic field applied perpendicular to the layers \cite{zhon18}.  In this paper, we investigate the spin excitations in similar crystals.

There are a number of commonly observed responses in the cuprates.
In the insulating parent compounds, antiferromagnetic spin waves disperse from magnetic Bragg peaks characterized by wave vector $\mathbf{Q}_{\rm AF}=(0.5,0.5,0)$ \cite{tran07}. With increasing hole-doping, the dispersion of the magnetic excitations tends to transform into an ``hourglass'' shape with incommensurate excitations at low energies, as has been observed in La$_{2-x}$Ba$_{x}$CuO$_{4}$ (LBCO), La$_{2-x}$Sr$_{x}$CuO$_{4}$ (LSCO), YBa$_{2}$Cu$_{3}$O$_{6+\delta}$ (YBCO), and Bi$_{2}$Sr$_{2}$CaCu$_{2}$O$_{8+\delta}$ (BSCCO) \cite{fuji12a}.  In optimally-doped cuprates, a spin gap is present below $T_c$, and a putative resonance peak is typically observed above the gap \cite{woo06,fong99,lake99,chan16c}.  The spin gap appears to be a common feature of cuprates with a spatially-uniform superconducting state \cite{li18}.  

With sufficient underdoping, however, there are several cases where the spin gap and resonance are not observed; instead, gapless spin fluctuations coexist with superconductivity.  Examples include YBa$_2$Cu$_3$O$_{6.45}$ with $T_c=35$~K \cite{hink08}, La$_{1.93}$Sr$_{0.07}$CuO$_{4}$ with $T_c=20$~K \cite{jaco15}, and La$_{1.905}$Ba$_{0.095}$CuO$_{4}$ with $T_c=32$~K \cite{xu14}.  In the latter case, decoupling of the superconducting layers has been observed over a considerable range of $c$-axis magnetic field \cite{wen12,steg13}.  The field-induced state is similar to the two-dimensional superconductivity detected at zero field in La$_{1.875}$Ba$_{0.125}$CuO$_4$ \cite{li07}, which has been attributed to pair-density-wave (PDW) superconducting order \cite{hime02,berg07}.  Is a similar story relevant to La-2126?

\begin{table*}[t]
\caption{\label{tab:data} Sample labels, superconducting transition temperatures $T_c$, along with incident neutron energies $E_i$, and corresponding $T_0$ and $F_1$ chopper frequencies used for the low energy (LE) and high energy (HE) configurations.  {\newr Also included are the energy resolution $\Delta E$ (full-width half-maximum at $\hbar\omega=0$) and $\Delta E/E_i$ \cite{arcs12}.}}
\begin{ruledtabular}
\begin{tabular}{cccccccccccccc}
 & & \ \ & \multicolumn{5}{c}{LE} & \ \ & \multicolumn{5}{c}{HE} \\
Sample & $T_c$ & & $E_i$ & $T_0$ & $F_1$ & $\Delta E$ & $\Delta E/E_i$ & & $E_i$ & $T_0$ & $F_1$ & $\Delta E$ & $\Delta E/E_i$ \\ 
 & (K) & & (meV) & (Hz) & (Hz) & (meV) & & & (meV) & (Hz) & (Hz) & (meV) &  \\
\hline
NSC & \ 0  & & 60 & 90 & 180 & 3.6 & 0.060 & & 150 & 90 & 300 & 7.9 & 0.053 \\
SC45 & 45 & & 40 & 30 & 180 & 2.1 & 0.052 & & 120 & 60 & 240 & 7.5 & 0.062\\
SC55 & 55 & & 60 & 90 & 180 & 3.6 & 0.060 & & 150 & 90 & 300 & 7.9 & 0.053 \\
\end{tabular}
\end{ruledtabular}
\end{table*}

Previous neutron-scattering measurements on La-2126 have focused on the elastic scattering of unannealed crystals that were non- or weakly superconducting \cite{ulri02,huck05}. These measurements showed long-range antiferromagnetic (AF) order in La$_{1.9}$Ca$_{1.1}$Cu$_{2}$O$_{6+\delta}$ and short-range AF order in La$_{1.85}$Sr$_{0.15}$CaCu$_{2}$O$_{6+\delta}$, with the order appearing to be commensurate in both cases \cite{huck05}. A tetragonal-to-orthorhombic structural phase transition upon cooling has been observed for both La$_{2-x}$Sr$_{x}$CaCu$_{2}$O$_{6+\delta}$ and La$_{2-x}$Ca$_{1+x}$Cu$_{2}$O$_{6+\delta}$ \cite{ulri02,huck05}. 

In this paper, we present an inelastic neutron scattering study of  La$_{2-x}$Ca$_{1+x}$Cu$_{2}$O$_{6}$ single crystals, two of which exhibit bulk superconductivity induced by high-pressure oxygen annealing.  We have previously characterized the structures of these crystals \cite{hu14b,schn17}.  The annealing appears to enhance the Ca concentration of the La-2126 phase into the superconducting regime, at the cost of creating some effective intergrowths of two other cuprate phases, as we explain below.  Nevertheless, only the La-2126 phase has bilayers, whose spin excitations have a modulated structure factor that allows us to separate them from other contributions.  Our key findings are 1) the absence of a spin gap in the superconducting state, and 2) a robust superexchange energy, despite considerable softening with doping.

The rest of the paper is organized as follows.  The materials and methods are described in the next section, followed by some background on the acoustic and optical magnon modes expected in a bilayer cuprate.  The data and analysis are presented in Sec.~IV.  We discuss the connection with intertwined orders in Sec.~V, and end with a summary.

\section{Materials and Methods}
\label{sec:MaterialsAndMethods}

The details of the growth, annealing, and characterization of crystals of \lcco\ with nominal values of $x=0.1$ and 0.15 are described in \cite{schn17,hu14b,zhon18}.  In the present study, two crystals of $x=0.1$ were studied: sample NSC (mass of 7.5~g) was as-grown and nonsuperconducting, while sample SC45 (6.3~g) was annealed in high-pressure oxygen (pressure above 0.55 GPa and $T>1130$~\degrees) and had a superconducting transition temperature $T_c = 45$~K (determined by magnetization).  Crystal SC55 (7.4~g) corresponds to nominal $x=0.15$, with similar annealing conditions, and $T_c=55$~K.

While the NSC sample appears to be single phase, detailed electron \cite{hu14b} and neutron \cite{schn17} diffraction studies demonstrated that the annealing induces the formation of intergrowth-like domains of \lco\ and La$_8$Cu$_8$O$_{20}$.  The volume fraction of the superconducting La-2126 phase was estimated to be $\sim70$\%\ \cite{schn17}.  It was inferred that the segregation of the \lco\ and La$_8$Cu$_8$O$_{20}$ phases caused an enrichment of the Ca concentration in the remaining La-2126 phase, so that the hole concentration in the CuO$_2$ bilayers is tuned by Ca substitution for La rather than by interstitial oxygen.   While it would be desirable to determine the actual Ca concentration in the La-2126 superconducting phase, doing so is complicated by the unusual heterostructures of these samples.

Inelastic neutron scattering experiments were performed on the SEQUOIA time-of-flight spectrometer at the Spallation Neutron Source, Oak Ridge National Laboratory \cite{sequoia10,tof_sns14}.  For each sample, two incident energies $E_i$ were used, a low-energy choice and a high-energy choice.  These are listed in Table~\ref{tab:data}, along with the corresponding settings for the $T_0$ and Fermi ($F_1$) chopper frequencies.  The different $E_i$ choices for SC45 vs.\ NSC and SC55 were made in separate experimental runs.  For simplicity, from here on we will refer to the LE and HE configurations as defined in the table.

To index the scattering, we use a pseudo-tetragonal unit cell with lattice parameters $a=3.83$ \AA\ and $c=19.36$~\AA.  All wave vectors are reported in reciprocal lattice units.  Each sample was oriented with ${\bf Q}=(H,0,L)$ in the horizontal plane and $(0,K,0)$ in the vertical direction.  For each measurement, the sample was rotated over at least $90^\circ$ in $1^\circ$ steps.  The collected data were then mapped into the region of $H \geq 0$, $K \geq 0$, and $L \geq 0$ by reflection across $H=0$, $K=0$ and $L=0$ and averaged (unless otherwise noted). Errorbars represent statistical error and correspond to 1 standard deviation from the observed value.  Data in the false-color intensity maps have been smoothed, and white regions indicate areas outside of the detector coverage.

For this paper, we will present the inelastic scattering data in the form of the dynamical scattering function $S({\bf Q},\omega)$ defined by the quantity $\tilde{M}({\bf Q},\omega)$  in Eq.~(10) of Ref.~\cite{xu13}, with the exceptions that we will not correct for the magnetic form factor (and the Debye-Waller factor is ignored).  {\newr A common way to determine the absolute magnitude of $S({\bf Q},\omega)$ is to normalize to measurements of incoherent elastic scattering from a standard vanadium sample.  As we only had vanadium data for the measurement conditions used for the SC45 sample, we used the following normalization procedure.}  First, for the SC45 crystal with the LE configuration and $T = 4$ K, the neutron scattering intensity of the (200) longitudinal acoustic phonon at $\hbar \omega = 6$ meV was measured and used to calculate the quantity $N R_0 k_f$, where $N$ is the number of Cu ions, $R_0$ is the resolution volume, and $k_f$ is the outgoing neutron wave vector magnitude \cite{xu13}. [Similar values of $N R_0 k_f$ (within $\sim$20\%) were obtained using  vanadium data.] Next, intensities of the NSC and SC55 data in the LE configuration were scaled relative to the SC45 data using acoustic phonon intensities at $(107)$ and $T=4$ K. As it was not possible to resolve the separate phonon branches near $(107)$, all of the spectral weight within $5.5 \leq \hbar \omega < 6.5$ meV, $0 \leq K < 0.2$, and $6.5 \leq L < 7.5$ was averaged and integrated along $H$ after subtracting a background obtained from fitting a sloped line to surrounding data. Finally, the HE-configuration data were normalized for each crystal relative to their LE data by comparing averages of incoherent scattering found at $(0.15,0.825,0)$, $(0.25,0.825,0)$, $(0.35,0.825,0)$, $(0.35,0.75,0)$, and $(0.5,0.15,0)$ at $T=4$~K. 

While we have not corrected for the magnetic form factor in evaluating $S({\bf Q},\omega)$, we will correct for it when converting to the imaginary part of the dynamic spin susceptibility, $\chi''({\bf Q},\omega)$.  As a result, the effective relationship used here is:
\begin{equation}
  \chi''({\bf Q},\omega) = \left[{\pi (1-e^{-\hbar\omega/kT}) \over 2f^2({\bf Q})}\right] S({\bf Q},\omega),
\end{equation}
where $f({\bf Q})$ is the magnetic form factor calculated for La$_2$CuO$_4$ in the same fashion as that for Sr$_2$CuO$_3$ analyzed in \cite{walt09}.  This formula essentially corresponds with Eq.~(13) of Ref.~\cite{xu13}.

\section{Background}

For small $x$, La$_{2-x}$(Ca,Sr)$_x$CaCu$_2$O$_6$ exhibits antiferromagnetic order with ordering wave vector ${\bf Q}_{\rm AF}=(0.5,0.5,1)$ \cite{ulri02,huck05}.  This involves antiparallel nearest-neighbor Cu spins in all directions within CuO$_2$ bilayers, similar to the order in \ybco\ \cite{tran88a}.  The coupling between neighboring layers causes the bilayer spin waves to split into acoustic and optical modes, where the oscillations in neighboring layers are correspondingly in-phase and anti-phase \cite{sato88}.  These modes have structure factors proportional to $\sin(\pi z'L)$ and $\cos(\pi z'L)$, respectively, where $z'c$ is the separation between Cu atoms along the $c$ axis, $z'=0.170$ \cite{kino92a}, and the scattered intensity is proportional to the square of the structure factor.   For the acoustic mode, the intensity is zero at $L = 0$ and 5.9, but a maximum at $L \approx 2.9$ \cite{huck05}; with the minima and maxima reversing for the optical mode.  The optical magnon mode should have a significant energy gap; in antiferromagnetic \ybco, the gap is $\sim70$~meV \cite{hayd96b,rezn96}, and it decreases in energy with doping \cite{pail06}.

Now, the bilayer spin fluctuations from the La-2126 phase are not the only source of scattering that may occur along ${\bf Q}=(0.5,0.5,L)$.  For example, the apical oxygens tend to be modulated at the same in-plane wave vector \cite{ulri02}, and we also have, in SC45 and SC55, a contribution from the antiferromagnetic correlations in the La-214 phase \cite{schn17}.   Spin fluctuations from the La-214 phase should vary monotonically with $L$, while the structure factor for the La-2126 apical-oxygen modulation is weak at small $Q$ and distinct from that of the bilayer spin fluctuations.  Hence, in order to identify the latter, it is important to measure the ${\bf Q}$ dependence of the spin fluctuations in three dimensions, which one can do with the rotating crystal method \cite{horace16}.  At lower energies, we can identify the La-2126 spin fluctuations from the $\sin^2(\pi z'L)$ intensity modulation, and at higher energies, we can use the loss of the modulation as a signature of the optical magnon gap.

\section{Data and Analysis}

\begin{figure}[t]
\begin{center}
\includegraphics[width=\columnwidth]{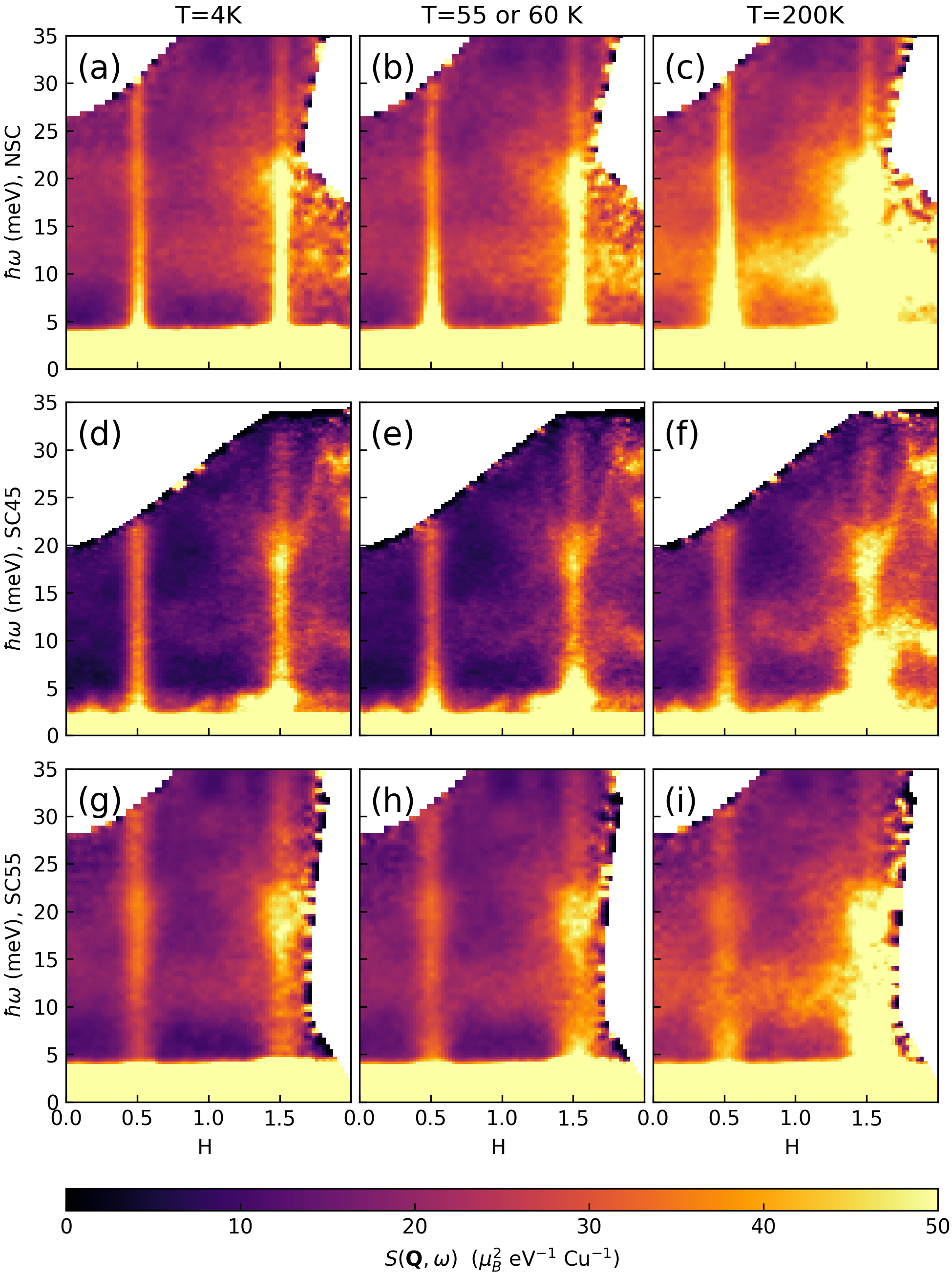}
\caption{\label{fig:EHSlices} Color maps of $S({\bf Q},\omega)$ obtained in the LE configuration as a function of excitation energy and ${\bf Q}=(H,0.5,3)$ for NSC (a-c), SC45 (d-f), and SC55 (g-i) at temperatures of 4~K (a,d,g) and 200~K (c,f,i) for all samples, 55~K for SC45 (e), and 60~K for NSC (b) and SC55 (h).  Data have been averaged within $2 \leq L \leq 4$ and $0.4 \leq K \leq 0.6$.  One can see the magnetic excitations dispersing steeply from $H=0.5$ and 1.5.
}
\end{center}
\end{figure}

\begin{figure}[t]
\begin{center}
\includegraphics[width=\columnwidth]{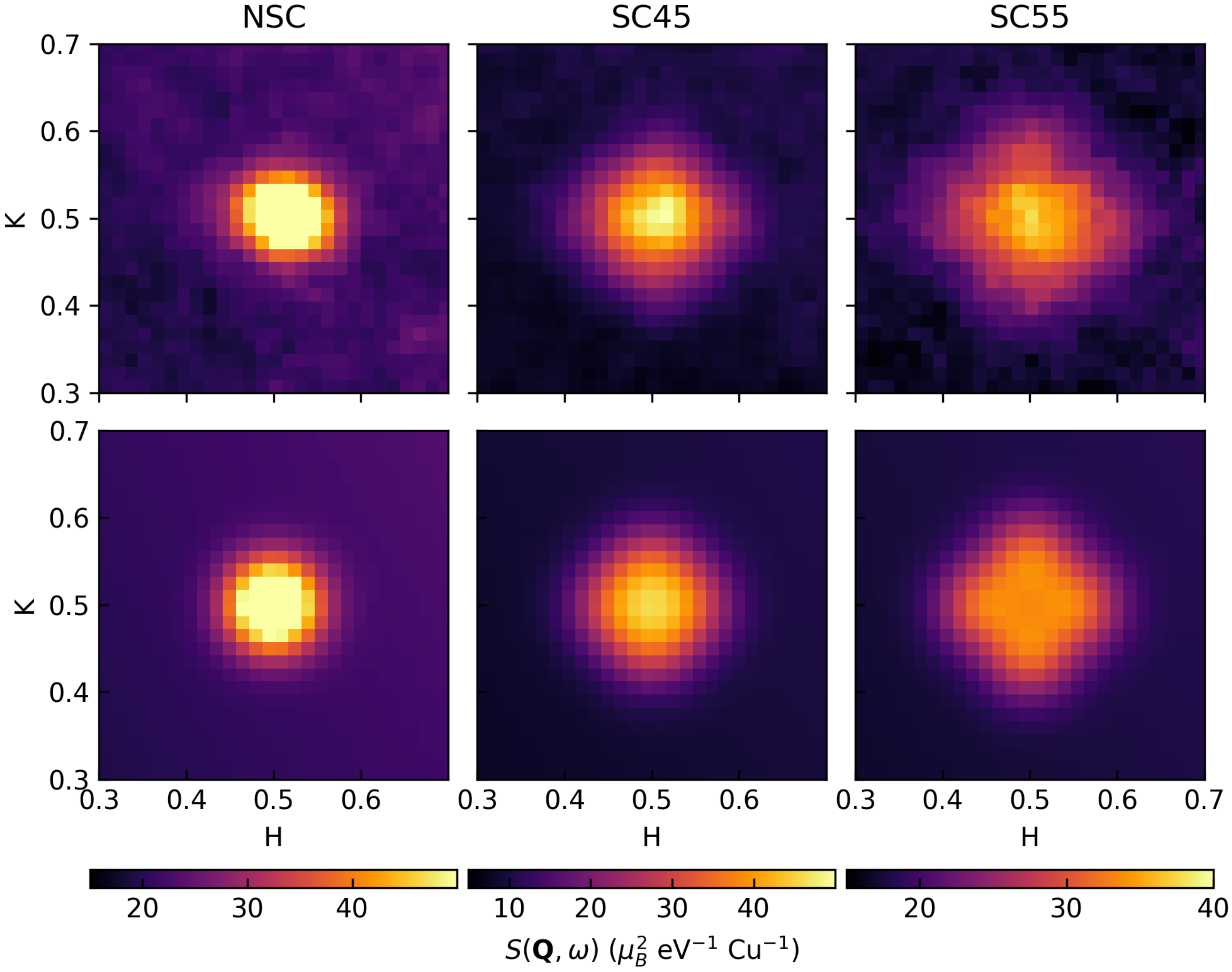}
\caption{\label{fig:diamondHKPlots} Top row: Color maps of $S({\bf Q},\omega)$ in the $(H,K,3)$ plane centered on the magnetic excitations at $(0.5,0.5,3)$. Data have been averaged within $2 \leq L \leq 4$ and $8 \leq \hbar \omega \leq 16$ meV. All data were taken at 4 K, with the LE configuration.  Bottom row: fits with 4 incommensurate gaussian peaks split about $(0.5,0.5)$, assuming FWHM~$ = 0.083$~rlu (obtained from fitting a single gaussian to the NSC result).  The incommensurabilities are 0.007(5), 0.044(1), and 0.056(1) for NSC, SC45, and SC55, respectively.
}
\end{center}
\end{figure}

Figure~\ref{fig:EHSlices} shows the dispersion of magnetic excitations along ${\bf Q}=(H,0.5,3)$ at low energies for all three samples.  Taking account of the expected acoustic mode structure factor for the CuO$_2$ bilayers (discussed in the previous section), the intensity has been averaged over the range $2\le L\le4$.  The dispersion about $H=0.5$ and 1.5 is unresolved because of the large effective spin-wave velocity.  With increasing $H$, superimposed phonon scattering grows in intensity.  Of particular note is the dispersive optical phonon mode with a minimum energy of 18 meV at $H=1.5$, with a strong resemblance to the feature reported in lightly-doped \lbco\ \cite{wagm15}.  

\begin{figure}[t]
\begin{center}
\includegraphics[width=\columnwidth]{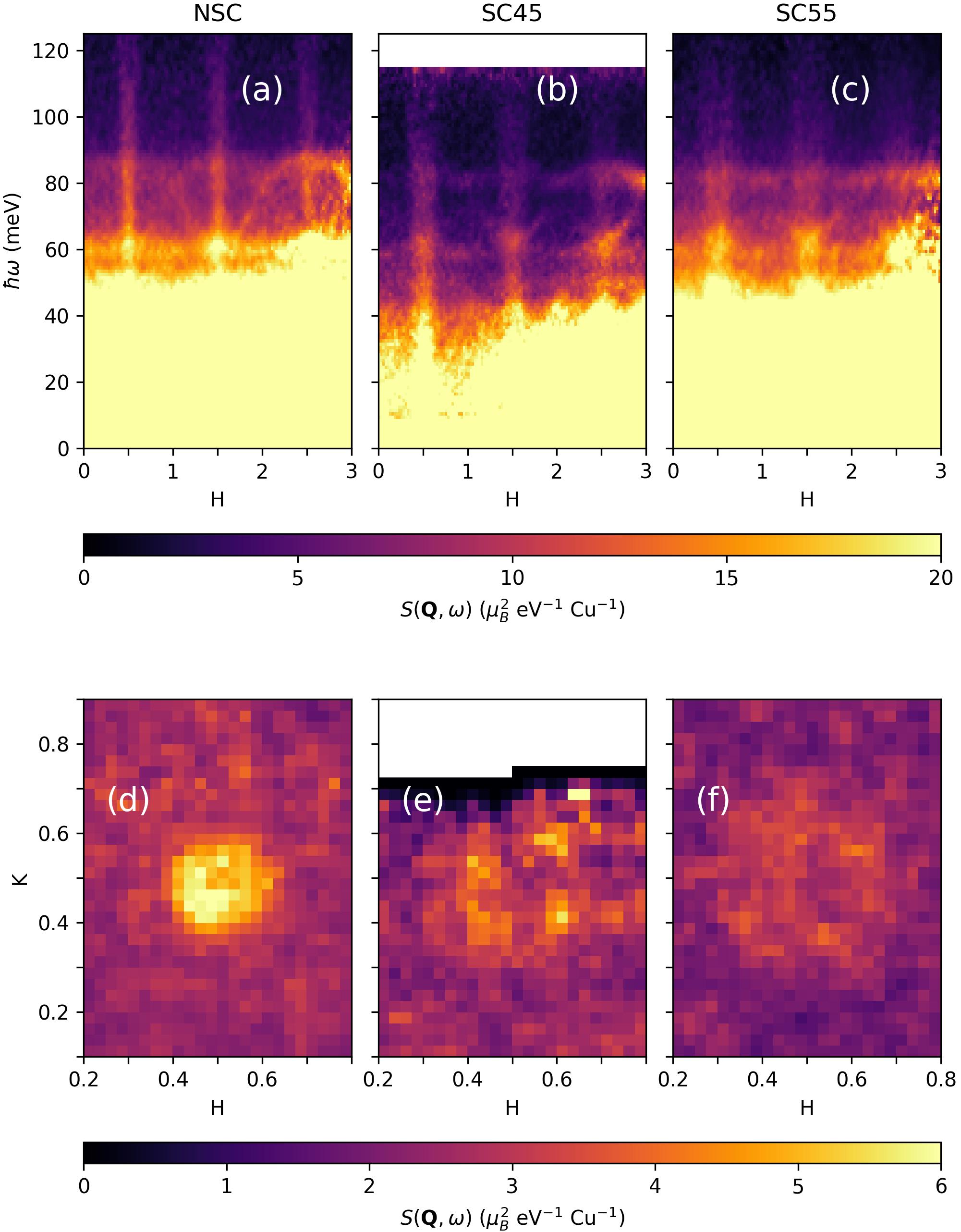}
\caption{\label{fig:EH120meVFigure}  Color maps of $S({\bf Q},\omega)$ obtained in the HE configuration at $T=4$~K along ${\bf Q} = (H,0.5,\langle L\rangle)$ for (a) NSC, (b) SC45, (c) SC55.  Data averaged within $0.4 \leq K \leq 0.6$ and within the full range of $L$ available for each $H$ and $\hbar \omega$ value.
(d-f) Corresponding constant-energy slices in the $(H,K,\langle L\rangle)$ plane at $\hbar \omega = 110$ meV. Data averaged within $100 \leq \hbar \omega \leq 120$ meV and the full range of available $L$ values (roughly $14 \leq L \leq 19$ for SC45 and $11 \leq L \leq 21$ for NSC and SC55). 
}
\end{center}
\end{figure}

To view the in-plane spin correlations, Fig.~\ref{fig:diamondHKPlots} shows intensity slices in the $(H,K,3)$ plane, where the energy has been averaged over the range of 8 to 16~meV.  (Note that we have averaged the data to improve statistics, and that nothing different from the average is apparent in any slices over smaller energy ranges.)  For the NSC sample, the signal is relatively round and narrow.  For the superconducting samples, the scattering is broader and adopts a diamond shape.  While no incommensurate structure is resolved, the orientation of the diamond is consistent with the incommensurability commonly observed in sufficiently-underdoped cuprate superconductors \cite{enok13}. The lower row of Fig.~\ref{fig:diamondHKPlots} shows fits to the data with 4 incommensurate peaks, with peak width contrained to match that of the NSC data.  {\newr (Since we cannot actually resolve incommensurate features in the data, a more elaborate analysis does not seem justified.)} The pattern found is similar to that observed in underdoped and twinned \ybco\ \cite{tran92}; in detwinned \ybco, the incommensurate splitting is almost resolvable at 3 meV \cite{hink08}.  The incommensurabilities fit to our superconducting samples are comparable to those for \ybco\ with $p\sim0.08$--0.09 \cite{enok13,haug10,dai01}.

\begin{figure}[t]
\begin{center}
\includegraphics[width=\columnwidth]{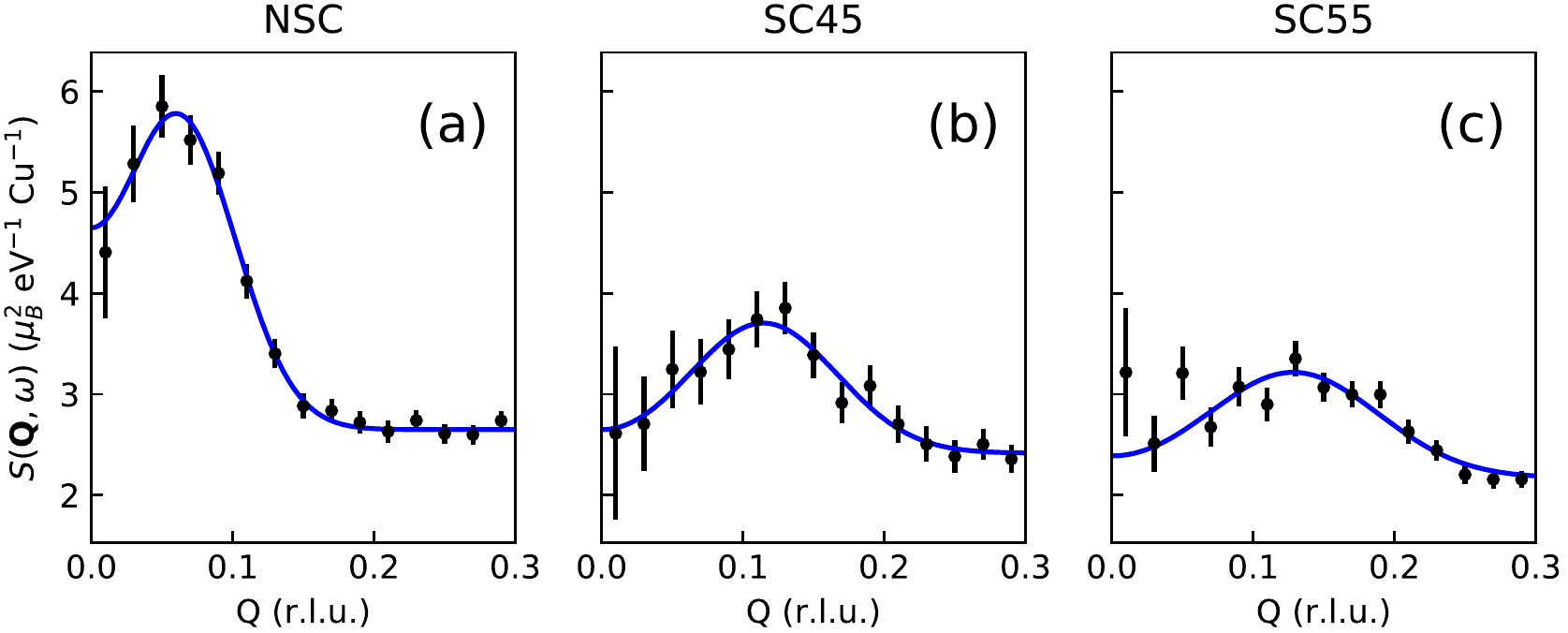}
\caption{\label{fig:radial}  (a)-(c) Radial average of ${\bf q}=(H,K,\langle L\rangle)-(0.5,0.5,\langle L\rangle)$ for the 110-meV data in Fig.~\ref{fig:EH120meVFigure}(d)-(f), respectively.  Circles with error bars: data; blue line: gaussian fit.  For the fits, a pair of identical peaks at $\pm Q_0$ were used.
}
\end{center}
\end{figure}

The magnetic dispersion to higher energies is shown in Fig.~\ref{fig:EH120meVFigure}(a)-(c).  Optical phonons extend up to $\sim85$~meV, but the excitations above that are purely magnetic.   The higher-energy magnetic excitations are readily apparent in NSC; they are weaker for SC45 and SC55 because of increased $Q$ width and damping from the doped holes, but one can see the roughly vertical dispersion of these excitations at lower energies, where they overlap with phonons.  Constant-energy slices at $\hbar\omega=110$~meV (averaged over a 20-meV window) are presented in Fig.~\ref{fig:EH120meVFigure}(d)-(f).  For NSC, we observe a ring, representing a cut through the conical dispersion of spin waves, whose slope is proportional to the in-plane superexchange energy $J$.  For SC45 and SC55, the ring has expanded, indicating a decrease in the effective $J$ consistent with previous cuprate results \cite{suga03,huck08}.   Figure~\ref{fig:radial} shows radial averages of the rings.  The gaussian fits yield ${\bf q}_0= 0.061(3)$, 0.115(8), and 0.128(6) rlu for NSC, SC45, and SC55, respectively, with half-widths of 0.048, 0.062, and 0.071 rlu.  From the ratio (110~meV)$/Q_0$ we can get an estimate of the spin-wave velocity and the effective $J$.  We will discuss this evaluation later, as we must take account of both the acoustic and optic fluctuations.  For now, let us denote the effective $J$ by the $T_c$ of the sample; then we note that $J_{45}/J_0= 0.53(5)$ and $J_{55}/J_0= 0.48(3)$.

\begin{figure}[t]
\begin{center}
\includegraphics[width=\columnwidth]{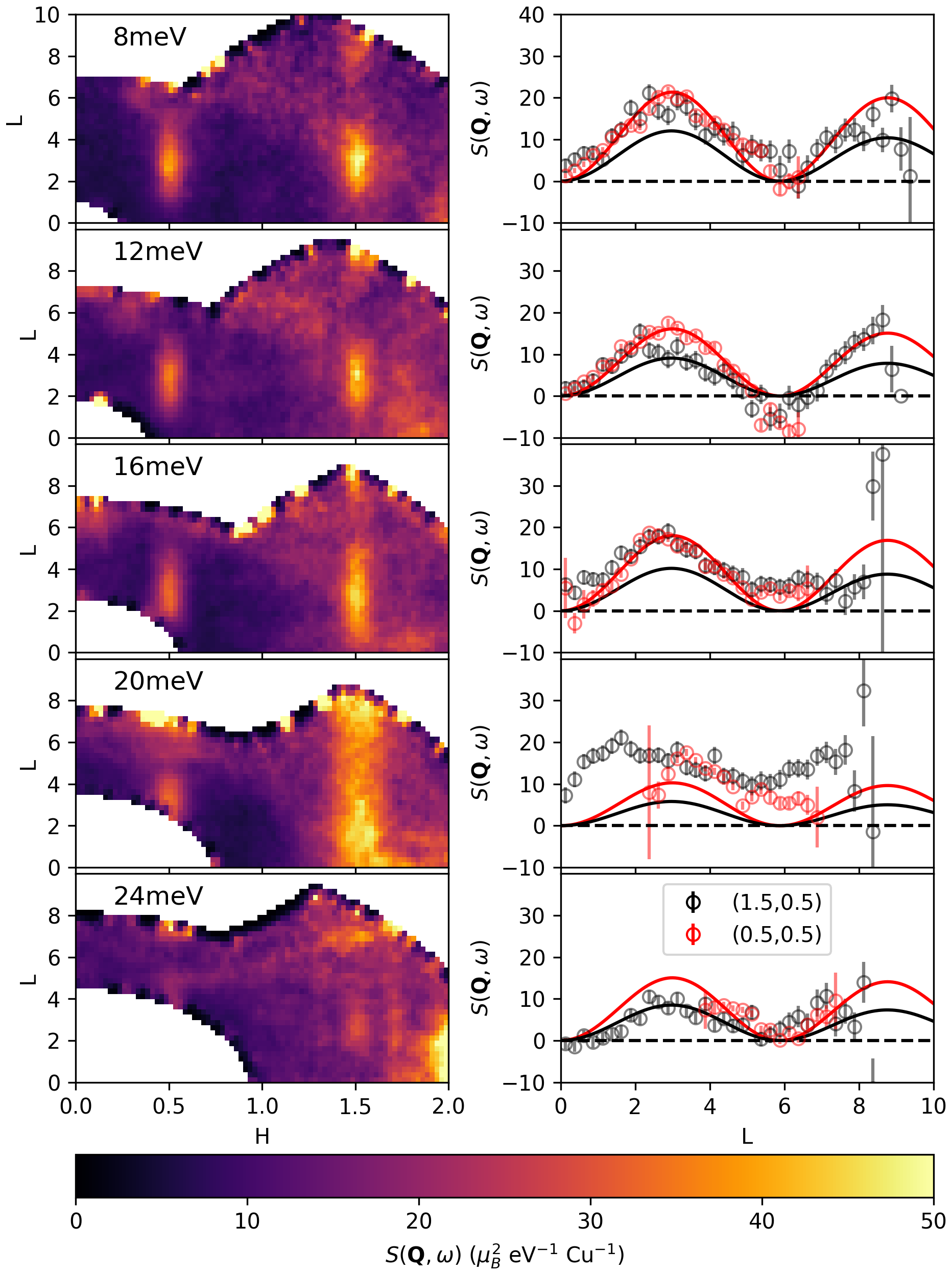}
\caption{\label{fig:HLSlicesAndCuts}  $S({\bf Q},\omega)$ for SC45, showing the magnetic scattering along $(0.5,0.5,L)$ and $(1.5,0.5,L)$. Left column shows constant-energy maps in the $(H,0.5,L)$ plane, averaged within $0.4 \leq K \leq 0.6$ and within $\Delta (\hbar \omega) = \pm 1$ meV around the energy transfers noted at the upped left of each panel  (8 to 24 meV in 4-meV steps). Data were taken at $T=4$ K and $E_i=40$ meV. Right column shows corresponding intensities of the magnetic excitations along $L$, averaged within $0.4 \leq K \leq 0.6$, $0.4 \leq H \leq 0.6$ for red circles and $1.4 \leq H \leq 1.6$ for black circles, after subtraction of background determined at neighboring wave vectors.  Solid lines correspond to $f^2({\bf Q})\sin^2(\pi z'L)$ using the Cu form factor determined in \cite{walt09}.
}
\end{center}
\end{figure}

\begin{figure}[t]
\begin{center}
\includegraphics[width=\columnwidth]{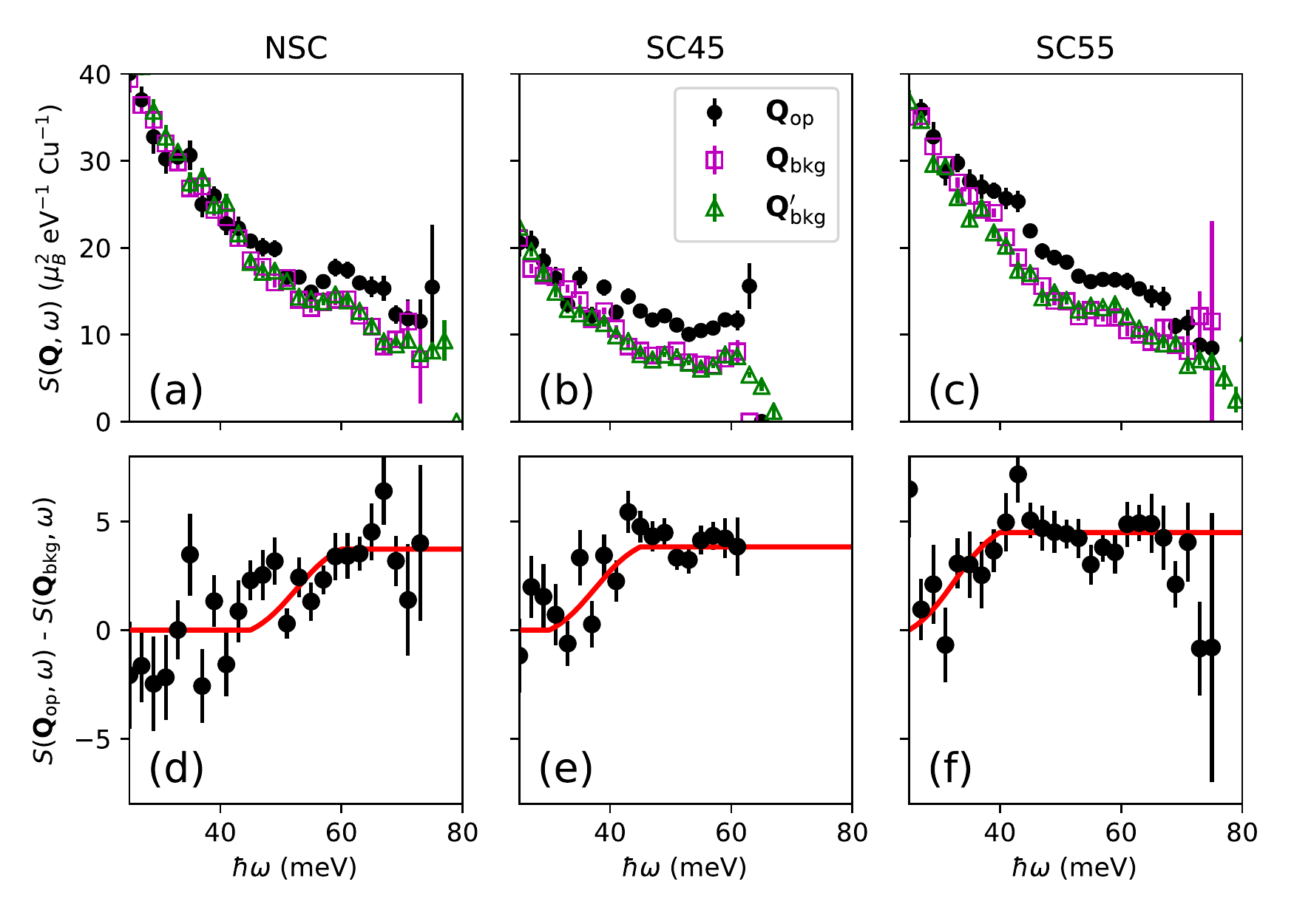}
\caption{\label{fig:OpticModeLDependence}  (a,b,c) $S({\bf Q},\omega)$ vs.\ $\hbar \omega$ at ${\bf Q}_{\rm op}=(0.5,0.5,6)$ (filled black circles), along with background intensity at ${\bf Q}_{\rm bkg}=(0.3,0.5,6)$ (open magenta squares) and ${\bf Q}'_{\rm bkg}=(0.7,0.5,6)$ (open green triangles). Data were averaged within $0.4 \leq H \leq 0.6$, $5 \leq L \leq 7$, and either $0.4 \leq K \leq 0.6$ for ${\bf Q}_{\rm op}$ or $0.2 \leq K \leq 0.8$ for the background points. (d,e,f) $S({\bf Q}_{\rm op},\omega)$ after subtraction of averaged background. Red lines are guides to the eye to highlight the increase in intensity with $\hbar\omega$, which is expected due to the presence of the optical branch. Measurements were done with the HE configuration at $T= 4$~K.
}
\end{center}
\end{figure}

To verify the acoustic magnon structure factor anticipated at low energies, we plot on the lefthand side of Fig.~\ref{fig:HLSlicesAndCuts} representative constant-energy slices of intensity in the $(H,0.5,L)$ plane;  cuts along $H=0.5$ and 1.5 are presented on the righthand side.  The data presented are for SC45 at $T=4$~K, but the results for the other samples and at temperatures up to 200 K show similar behavior.  For $\hbar\omega=8$, 12, 16, and 24 meV, one can see that the scattered intensity is sinusoidally modulated, with a maximum at $L\approx3$ and minima at $L=0$ and 6, as expected.  The only exception is at $\hbar\omega=20$~meV, where overlapping scattering from an optical phonon dominates the response.  We also note that, for $\hbar\omega=8$ and 16 meV, the intensities at $H=1.5$ are closer to those at $H=0.5$ than expected for the form factor dependence.

When the optical magnon branch appears at higher energies, it should have an intensity maximum at $L=6$, where the acoustic structure factor is zero.  To look for the onset of the optical mode, we plot the intensity at ${\bf Q}_{\rm op}=(0.5,0.5,6)$ and at a background point ${\bf Q}_{\rm b}$ as a function of $\hbar\omega$ for each sample in Fig.~\ref{fig:OpticModeLDependence}(a)-(c).  Consistent with expectations, we see that the signal at ${\bf Q}_{\rm op}$ matches that at ${\bf Q}_{\rm b}$ for low energies but rises above background at higher energies.  To see this more clearly, we plot the difference in intensities at ${\bf Q}_{\rm op}$ and ${\bf Q}_{\rm b}$ in Fig.~\ref{fig:OpticModeLDependence}(d)-(f).  From these results, we estimate an optical magnon gap   $\Delta_{\rm op}$ of $53\pm5$~meV, $38\pm5$~meV, and $33\pm5$~meV for NSC, SC45, and SC55, respectively.  

For the case of undoped La-2126, where a spin-wave model similar to that for YBa$_2$Cu$_3$O$_6$ \cite{tran89} should be appropriate, the optical magnon gap, $\Delta_{\rm op}$, should be proportional to $\sqrt{J_\perp J}$, where $J_\perp$ is the effective nearest-neighbor interlayer exchange within bilayers.  Taking the square of the ratio of the optical gaps, we find 
\begin{equation}
{ \Delta^2_{\rm op}({\rm SC45}) \over \Delta^2_{\rm op}({\rm NSC}) } = 0.51(13) = {J_{45\perp}J_{45}\over J_{0\perp}J_0 }\approx {J_{45}\over J_0},
\end{equation}
and
\begin{equation}
{ \Delta^2_{\rm op}({\rm SC55}) \over \Delta^2_{\rm op}({\rm NSC}) } = 0.39(12) = {J_{55\perp}J_{55}\over J_{0\perp}J_0 }\approx {J_{55}\over J_0}.
\end{equation}
From this analysis, it appears that the doping dependence of $\Delta_{\rm op}$ is fully accounted for by the change in $J$, suggesting that $J_\perp$ is not significantly affected by doping.

Now that we have estimates for $\Delta_{\rm op}$, we can evaluate $J$ from the fits in Fig.~\ref{fig:radial}.  We assume that the acoustic branch disperses linearly in $q$ with velocity $v$.  The dispersion of the optical mode is assumed to be $\sqrt{\Delta^2_{\rm op} + \hbar^2v^2q^2}$.  For the measurements at 110~meV, both modes should be contributing approximately equally.  If $\bar{q}$ is the average peak position, then the velocity should be given by
\begin{equation}
  v = {\omega\over 2\bar{q}}\left[1+\sqrt{1-(\Delta_{\rm op}/\hbar\omega)^2}\right].
\end{equation}
The theoretical spin-wave velocity for the nearest-neighbor Heisenberg antiferromagnet appropriate to a CuO$_2$ plane is \cite{sing89}
\begin{equation}
  v = \sqrt{2}Z_cJa/\hbar,
\end{equation}
where $Z_c=1.18$.  From these formulas and the data for NSC, we obtain $J=161_{-46}^{+105}$~meV, where we have used half of the half-width at half maximum as an estimate of uncertainty in $\bar{q}$.  This value of $J$ (with its substantial uncertainty) is on the high-side of the distribution of results for other cuprates, but that distribution has a significant width \cite{tran07}.


\begin{figure}[t]
\begin{center}
\includegraphics[width=\columnwidth]{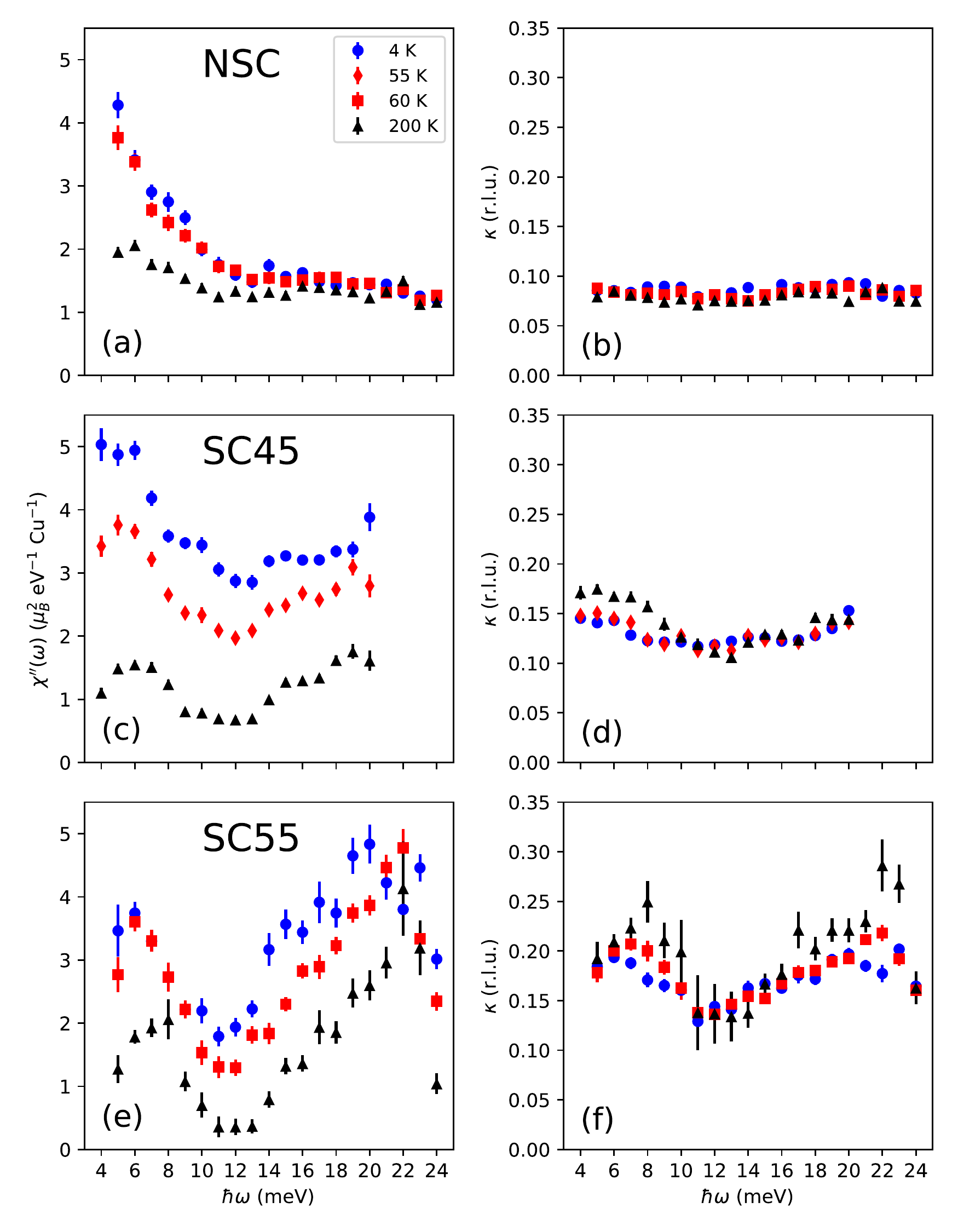}
\caption{\label{fig:constantQCut2DFits}  Fitting results for $\chi''(\omega)$ (a,c,e) and $\kappa$ (full $Q$ width at half maximum) (b,d,f) for NSC (a,b), SC45 (c,d), and SC55 (e,f), as described in the text. These results are from LE data; measurement temperatures are indicated in the legend.}
\end{center}
\end{figure}

To characterize the strength of the magnetic excitations as a function of temperature and energy transfer, we took the LE data, integrated it over $2\le L\le4$, divided the energy scale into 0.5-meV bins, and fit the scattered signal to the form
\begin{equation}
  S({\bf Q},\omega)/f^2({\bf Q}) = p_0 + {\bf p}_1\cdot {\bf Q} + {A\over 2\pi\sigma^2}e^{-\frac12({\bf Q}-{\bf Q}_{\bf AF})^2/\sigma^2},
\end{equation}
where the parameters $p_0$, ${\bf p}_1$, $A(\omega)$, and $\sigma(\omega)$ were fit for each $\hbar\omega$ bin, and the full-width half-maximum $\kappa = 2\sqrt{2\ln 2}\sigma$.  To convert the amplitude to the ${\bf Q}$-integrated imaginary part of the dynamic susceptibility, $\chi''(\omega)$, we used
\begin{equation}
 \chi''(\omega) = \frac{\pi}{2} A(\omega)(1-e^{-\hbar\omega/k_{\rm B}T}).
\end{equation}
We plot the results for $\chi''(\omega)$ and $\kappa(\omega)$ for each sample at several temperatures in Fig.~\ref{fig:constantQCut2DFits}.

For sample NSC, $\chi''(\omega)$ is constant in energy and temperature for $\hbar\gtrsim10$~meV, consistent with the response observed from spin waves \cite{sham93,hayd96b}.  At low $\hbar\omega$, the rise in the signal appears to be a quasielastic response, suggesting local hopping by the low density of doped holes frustrates static order.  The spectrum looks similar to that observed recently in La$_{1.93}$Sr$_{0.07}$CuO$_4$ \cite{jaco15}.  

\begin{figure}[b]
\begin{center}
\includegraphics[width=\columnwidth]{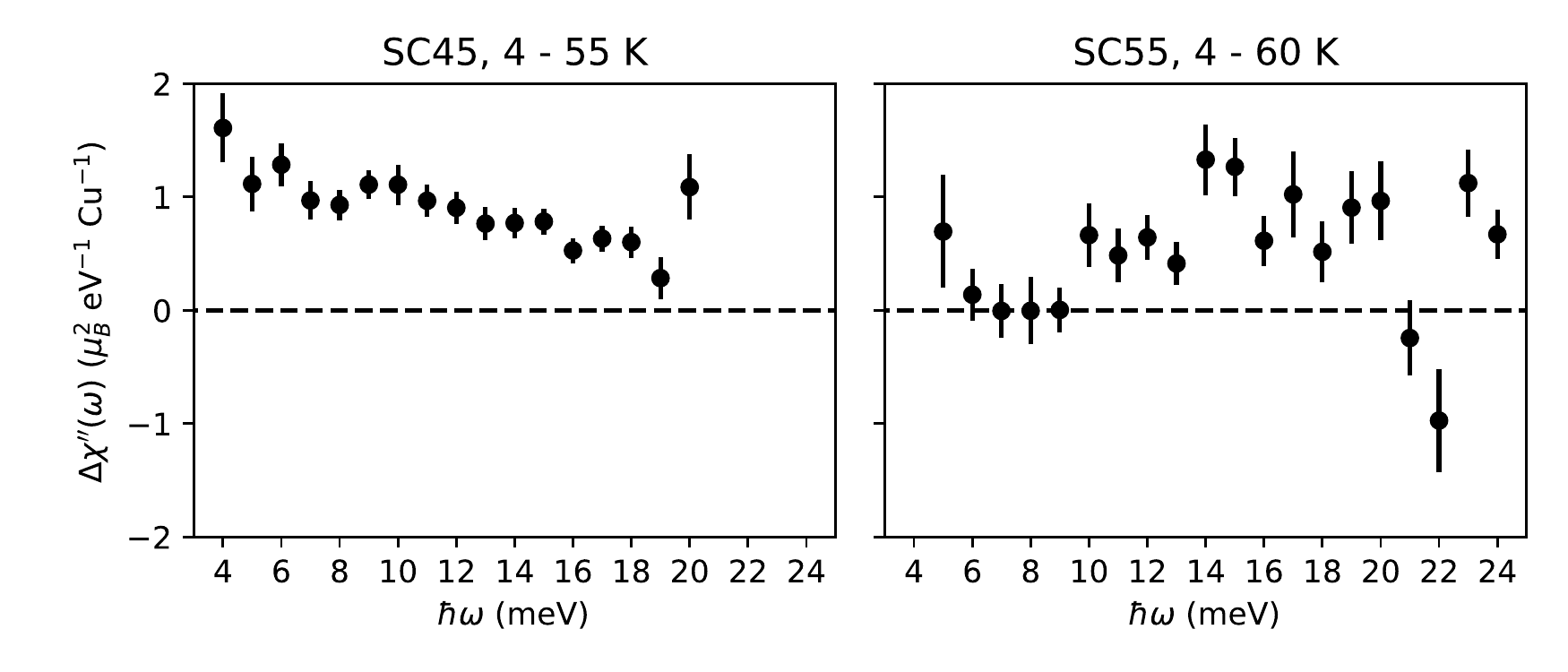}
\caption{\label{fig:diffPlot} Differences of $\chi^{\prime \prime} (\omega)$ across $T_{c}$, with $T=55$-K (60-K) data subtracted from 4-K data for SC45 (SC55) crystals.}
\end{center}
\end{figure}

For samples SC45 and SC55, the magnetic response is reduced at high temperature and grows on cooling, in a fashion qualitatively similar to that observed in other underdoped cuprate superconductors \cite{hink08,li08b,lips09,xu14,tran92}.  At low temperature, there is a rise in response at low frequency, which is different from the behavior in cuprates that exhibit a spin gap \cite{bour97,dai99,chri04,chan16c,xu09}.  To confirm the change across $T_c$, we plot the difference between 4 K and $T\gtrsim T_c$ in Fig.~\ref{fig:diffPlot}.  As one can see, there is no sign of a spin gap or resonance.  The coexistence of superconductivity with gapless spin fluctuations has been observed previously in underdoped \lbco\ \cite{xu14}, \lsco\ \cite{jaco15}, and very underdoped \ybco\ \cite{stoc08,hink08,li08b}.

The width parameter $\kappa$, plotted in Fig.~\ref{fig:constantQCut2DFits}(b), (d), and (f), grows with doping but does not show any significant dependence on temperature or energy over this modest energy range.  While we have already noted that we cannot resolve any incommensurability, the variation in $\kappa$ with increasing $T_c$ is consistent with the variation in incommensurability with doping seen in other cuprate superconductors \cite{enok13}.


\section{Discussion}

The features of the magnetic excitation spectra that we have observed in La-2126 crystals are similar to those observed in various other cuprate systems \cite{fuji12a}.  For example, the NSC sample exhibits a large spin velocity, indicating an effective $J$ that may be larger than the $J\approx 143$~meV of \lco\ \cite{head10}; measurements to significantly higher excitation energies would be necessary to reduce the uncertainty in $J$.   The decrease in the magnitude of $J$ with doping is consistent with 2-magnon Raman scattering results for other cuprates \cite{suga03}; the drop by 50\%\ relative to the insulating parent corresponds with optimal doping in \lsco\ and \bscco.  The decrease in spectral weight at 110 meV with doping follows the trend established by Stock {\it et al.} \cite{stoc10}: the magnetic spectral weight remains strong at low energies but is depressed on the scale of the pseudogap energy (corresponding to twice the antinodal pseudogap energy identified by photoemission) \cite{hufn08}.  

The low-energy spin excitations exhibit the bilayer acoustic-mode structure factor previously observed in \ybco\ \cite{tran89,pail04} and \bscco\ \cite{capo07}.  Our result of $\Delta_{\rm op}=53(5)$~meV can be compared with results for \ybco\ with $x=0.15$ where $\Delta_{\rm op}=74(5)$~meV and $J=125(5)$~meV were found \cite{hayd96b} (with slightly reduced values reported for a sample with $x=0.20$ \cite{rezn96}).  Our smaller $\Delta_{\rm op}$ and larger $J$ in the NSC sample indicate a reduced $J_\perp$ compared to YBCO.  On doping \ybco\ to $x=0.5$ ($T_c=52$~K), $\Delta_{\rm op}$ dropped to $\sim59$~meV \cite{bour97}, which goes in the same direction as our observations.

For energies below $\Delta_{\rm op}$, the acoustic-mode structure factor allows us to separate the magnetic excitations of La-2126 from a possible contribution from the La-214 intergrowth layers in the annealed crystals, as the latter response should be independent of $L$.  It follows that we have good confidence that our observations of gapless spin fluctuations in SC45 and SC55 correspond to intrinsic features.  We certainly see some evolution of the low-energy excitations with doping shown in Fig.~\ref{fig:constantQCut2DFits}, as $\chi''(\omega)$ for $\hbar\omega \lesssim 20$~meV is reduced in the superconducting samples at 200~K, with a pronounced dip at 12~meV.  Nevertheless, that weight grows on cooling to near $T_c$, and the difference across $T_c$ (Fig.~\ref{fig:diffPlot}) shows no dip.

There is also an enhancement of $\chi''(\omega)$ at $\sim20$~meV in the SC55 sample compared to the NSC sample.  This feature appears to be similar to the one identified previously in lightly-doped \lbco\ by Wagman {\it et al.} \cite{wagm15}, observed in other La-214 samples \cite{xu14,li18},  and attributed to spin-phonon hybridization \cite{wagm16}.   The apparent absence of such a feature in the NSC sample suggests that the coupling between spin-excitations and phonons requires a sufficient level of hole doping.  We note that a related gapped-magnon feature with $\hbar\omega\sim25$~meV was observed in stripe-ordered La$_{1.67}$Sr$_{0.33}$NiO$_4$ \cite{woo05} and confirmed to be magnetic with the use of polarized neutrons \cite{free08}.

The absence of a spin gap plus the magnetic-field-induced decoupling of superconducting layers \cite{zhon18} parallels such behavior in \lbco\ with $x=0.095$ \cite{xu14,wen12,steg13}.  Of course, there are some differences, as well.  In the case of LBCO, there is direct evidence for both spin and charge stripe order \cite{wen12,huck11}.  For the present samples, we cannot resolve incommensurate peaks in the spin fluctuations, although the data are compatible with broad incommensurate peaks.  This situation is similar to YBCO with $x=0.45$, where detwinned crystals show a clear nematic anisotropy in the magnetic peak widths at low temperature \cite{hink08}; in that system, incommensurability of the lowest-energy excitations becomes resolvable at higher doping \cite{bour00,hayd04}.  Another issue is the absence of evidence for charge order in La-2126 and YBCO with $x=0.45$.  The lack of evidence does not rule out charge order, but indicates one direction for further study.

As noted in the Introduction, the decoupling of superconducting layers in LBCO, both in zero field for $x=0.125$ \cite{li07} and in applied field for $x=0.095$ \cite{wen12,steg13}, has been attributed to PDW order associated with the stripe order \cite{hime02,berg07}.  So far, direct evidence for PDW order has been limited.  The best evidence so far comes from a scanning tunneling spectroscopy study \cite{edki18} of modulations in vortex halos in \bscco, where a specific charge modulation was detected that was predicted to occur when uniform and PDW superconducting orders coexist \cite{berg09b}.  For LBCO with $x=1/8$, phase-sensitive measurements in a Josephson junction provide evidence for a charge-$4e$ response, which is another predicted behavior associated with PDW order \cite{berg09c}.  A recent high-magnetic-field transport study on the same compound has provided evidence that the charge stripes contain pair correlations even in the absence of phase coherence between neighboring stripes \cite{li19}.

The nature of superconducting order in La-2126 deserves further attention.  While evidence for PDW order is quite indirect, the gapless spin fluctuation spectrum and its temperature dependence seem to indicate that some combination of intertwined orders \cite{frad15} is involved.

\section{Summary}

Three different single crystals of La$_{2-x}$Ca$_{1+x}$Cu$_{2}$O$_{6+\delta}$ for $x=0.10$ and $x=0.15$, two of which were made superconducting by high-pressure oxygen annealing, have been studied by inelastic neutron scattering. For the NSC sample, the temperature-dependence of the {\bf Q}-integrated $\chi''(\omega)$ was found to be very similar to that of weakly-hole-doped cuprates. The superconducting crystals, on the other hand, showed a reduction in $\chi^{\prime \prime}(\omega)$ with increasing temperature; however, the change in spectral weight across $T_c$ did not show any sign of the spin gap or resonance features seen in many other cuprate superconductors. The lack of these features is consistent with data from underdoped La-214 cuprates.   The coexistence of gapless spin fluctuations with superconductivity suggests that intertwined orders are involved.  Further work is necessary to test for possible PDW order.

\section*{Acknowledgments}

The work at Brookhaven was supported by the Office of Basic Energy Sciences, U.S. Department of Energy (DOE) under Contract No.\ DE-SC0012704. This research used resources at the  Spallation Neutron Source, a DOE Office of Science User Facility operated by the Oak Ridge National Laboratory.

\bibliography{LNO,theory,neutrons}

\end{document}